\documentclass[12pt]{iopart}

\usepackage{graphicx}% Include figure file
\usepackage{epsf}% Include figure file
\usepackage{epsfig}

\begin{document}
%%%%%%%%%%%%%%%%%%%%%%%%%%%%%%%%%%%%%%%%%%%%%%%%%%%%%%%%%%%%%%%%%%%%%%%%%%%%%%%
\title[Electronic structure and]{ Electronic structure and magnetism in doped semiconducting 
half-Heusler compounds} 
\author { B.R.K. Nanda and I. Dasgupta\footnote[3]{To whom correspondence should be addressed (dasgupta@phy.iitb.ac.in)}}
\address{Department of Physics, Indian Institute of Technology Bombay, Powai, Mumbai 400076, India}
%%%%%%%%%%%%%%%%%%%%%%%%%%%%%%%%%%%%%%%%%%%%%%%%%%%%%%%%%%%%%%%%%%%%%%%%%%%%%%%

\begin{abstract}
We have studied in details 
the electronic structure and magnetism in M (Mn and Cr)  
doped semiconducting half-Heusler compounds FeVSb, CoTiSb and 
NiTiSn (XM$_{x}$Y$_{1-x}$Z) in a wide concentration range using 
local-spin density functional method in the framework of 
tight-binding linearized muffin tin orbital method(TB-LMTO) and supercell approach.
Our calculations indicate that some of these compounds 
are not only ferromagnetic but also half-metallic and may be useful for 
spintronics applications. The electronic structure of the doped systems 
is analyzed with the aid of a simple model where we have considered the 
interaction between the dopant transition metal (M) and the valence band 
X-Z hybrid. We have shown that the strong X-d - M-d interaction places the M-d states close to the Fermi level with the M-t$_{2g}$ states 
lying higher in energy in comparison to the M-e$_{g}$ states. 
Depending on the number of available d-electrons, 
ferromagnetism is realized provided the d-manifold is partially 
occupied. The tendencies toward ferromagnetic(FM) or antiferromagnetic(AFM)
behavior are discussed within Anderson-Hasegawa models of super-exchange and 
double-exchange. In our calculations for Mn doped NiTiSn, the strong preference for FM over AFM ordering suggests a possible high Curie temperature for 
these systems. 
%\pacs{71.20.-b, 71.20.Be, 75.50.Cc, 75.50.Pp}

\end{abstract}
\maketitle

%%%%%%%%%%%%%%%%%%%%%%%%%%%%%%%%%%%%%%%%%%%%%%%%%%%%%%%%%%%%%%%%%%%%%%%%%%%%%
\section{Introduction}
Spin-based electronics or spintronics is currently an active area of
research because it provides a possibility to integrate electronic, opto-electronic and magneto-electronic multifuntionality on a single device exploiting
both spin as well as charge degree of freedom of an electron. The
half-metallic ferromagnets having only one electronic spin direction
at the Fermi energy resulting in 100$\%$ spin-polarization
and ferromagnetic semiconductors 
where magnetic and semiconducting properties can be controlled and tuned 
are suggested to be suitable candidates
for spintronic applications. The recent interest in ferromagnetic
semiconductors for spintronic applications is spurred
by the pioneering work by Ohno and co-workers\cite{Ohno}, in the late nineties, showing that a ferromagnetic
Curie temperature as high as 110 K can be achieved in Mn doped GaAs,
demonstrating the feasibility that ferromagnetic property can be incorporated
in traditional semiconductors. This work has stimulated vigorous experimental
as well as theoretical activity in diluted magnetic semiconductors (DMSs).
As a result, on the experimental front, ferromagnetism in
diluted magnetic semiconductors, in some cases with T$_{c}$ close to room
temperature have been reported for Mn doped GaP\cite{Ntheodor}, Mn-doped Chalcopyrite
CdGeP$_{2}$\cite{Medvedkin} and Mn doped GaN\cite{Ntheodor1,Mlreed}. Ferromagnetism has also been reported for
oxide based diluted magnetic semiconductors such as
Co doped TiO$_{2}$\cite{Matsumoto}, SnO$_{2}$\cite{Ogale} and La$_{1-x}$Sr$_{x}$TiO$_{3-\delta}$\cite{Zhao}.

The present theoretical understanding of ferromagnetism in DMSs 
is very limited. The widely accepted
model Hamiltonian\cite{Matsukura} for Mn-doped
DMSs {\it e.g.} Ga$_{1-x}$Mn$_{x}$Sb
considers the impurity state bound to Mn ion to be  shallow acceptors formed from the host material
and the interaction between the Mn impurity spin orientations to be mediated by valence band holes.
The exchange interaction between the localized transition metal impurities
and the hole is suggested to be RKKY like, and it has been argued that such a model can capture salient features of the diluted magnetic semiconductors. The other possible mechanism
that can stabilize ferromagnetism in DMS is suggested to be 
Zener's p-d exchange\cite{dietl}. On the other hand,
the first principles calculations suggest that the Mn
induced hole can have significant 3-d character. 
In the series \cite{Sato} Mn doped
GaN $\rightarrow$ GaP $\rightarrow$ GaAs $\rightarrow$
GaSb the hole generated by introducing Mn in GaN is found to have significant 3d character, while in GaSb the hole is found to have primarily host character.
Further the first principles calculations suggest that
ferromagnetism is stabilized by kinetic energy driven
mechanism either by hybridization induced negative exchange splitting \cite{ddsarma} or double
exchange\cite{Akai} depending on the nature and the location of the holes.

The discussion of the preceding paragraph suggests that the
mechanism behind ferromagnetism in DMSs is still far from complete.
It also raises the important questions
whether any transition metal (TM) -semiconductor combination will result in ferromagnetism and
what are the relevant energetics that govern the interaction of the
transition metal with the host resulting in ferromagnetism.
To understand some of these issues the current research effort
in spintronics materials is also directed toward
search for potential DMS hosts. Recently
it was found experimentally\cite{Uher} that
semiconducting tetradymite structure material
Sb$_{2}$Te$_{3}$ act as
DMS with T$_{c}$ $\simeq$ 20K when doped with
few per cent of V atoms. Similarly a recent,
first principles calculation, \cite{Lambrecht}
indicate that transition metal doped SiC may be ferromagnetic with T$_{c}$ very close to room temperature. 
In the present work we have considered semiconducting half-Heusler systems as a potential DMS host.

The half-Heusler compounds with the general formula XYZ, where
X and Y are transition metals and Z is a sp-valent element has been a subject
of continuous attention because of their variety of novel magnetic properties
ranging from antiferromagnets\cite{cumnsb}, half-metallic ferromagnets\cite{Groot1}, weak ferromagnets, Pauli metals, semimetals\cite{xia1} to semiconductors\cite{evers}.
In particular, the half-metallic semi-Heusler alloys can be 
attractive for spintronics applications due to their relatively high 
Curie temperature and similarity of the crystal structure
to zinc-blende structure which are adopted by a large
number of semiconductors like GaAs, ZnSe, InAs {\it etc.}.
In the present work, inspired by DMSs
we have examined in details the electronic structure of 18 valence electron semiconducting half-Heusler alloys FeVSb, CoTiSb and NiTiSn when doped with
transition metal impurities Mn and Cr. The electronic structure of 
Mn doped semiconducting 
half-Heusler alloys have also been investigated in the framework 
of KKR-CPA calculations\cite{Tobola}. 
The doped Heusler systems has also 
been a subject of interest due to the possibility
of realizing half-metallic antiferromagnetism\cite{Groot}
 Half-metallic antiferromagnets  are half-metallic systems
with vanishing macroscopic moment. 
Our calculations for the doped semi-conducting half-Heusler systems
suggests, for impurity concentrations as low as 3 \%
some semiconducting half-Heusler systems are transformed
to half-metallic ferromagnets, with possibly with high Curie temperature.

The rest of the paper is organized as follows. In section 2 we describe
the crystal structure and the computational details. A detailed analysis of the
electronic structure and magnetism
of doped Heusler systems is carried out in section 3.
Finally, in section 4 we present our conclusions.

\section{Structure and computational details}

%%%%%%%%%%%%%%%%%%%%%%%%%%%%%%%%%%%%%%%%%%%%%%%%%%%%%%%%%%%%%%%%%%%%%%%%%%
% A picture of structure should come here (Fig. 1)
\begin{figure}
\begin{center}
\epsfig{figure=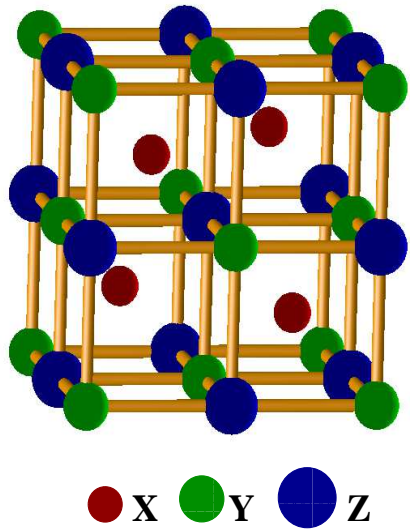,width=0.4\textwidth,height=0.3\textheight}
\caption{The cubic half-Heusler structure}
\end{center}
\end{figure}
%%%%%%%%%%%%%%%%%%%%%%%%%%%%%%%%%%%%%%%%%%%%%%%%%%%%%%%%%%%%%%%%%%%%%%%%%

The half-Heusler compounds XYZ crystallize in the face centered cubic structure
with one formula unit per unit cell as shown in figure 1. The space group is
F4/3m (No 216).  For the semiconducting half-Heusler systems 
considered here the higher valent transition elements
Fe,Co and Ni represent
X while the lower valent transition element 
V, Ti represent Y and Sb/Sn represent Z. In the
conventional stable structure Y and Z atoms are located at 4a(0,0,0) and
4b($\frac{1}{2}$,$\frac{1}{2}$,$\frac{1}{2}$) positions, forming the rock
salt structure arrangement. The X atom is located in the octahedral
coordinated pocket, at one of the cube center positions 4c($\frac{1}{4}$,
$\frac{1}{4}$,$\frac{1}{4}$) leaving the other 4d($\frac{3}{4}$,$\frac{3}{4}$,
$\frac{3}{4}$) empty. When the Z-atomic positions are empty
the structure is analogous to zinc blende structure, which is common for 
large number of semiconductors. 

In order to study the effect of 
Mn and Cr impurities (M) in semi-conducting 
half-Heusler systems ({\it i.e.} to simulate the effect of doping) 
we have constructed supercells with 
size dependent on the percent of M doping. In each supercell we have replaced
either one or two Y atoms by M atoms. For the two impurities 
embedded in the supercell we have studied the 
electronic structure of these doped systems both as a 
function of orientation of one M atom with respect to the other as well as 
distance between the two M atoms. Further embedding a
pair of impurity atoms also allowed us to study 
the interaction between a pair of M moments. 
The largest supercell chosen was 64 times 
the original unit cell and consisted of 192(256) atoms 
(including the empty spheres)
to simulate 3.125 $\%$ concentration for a pair of M atoms.
The size of the supercell 
was chosen to ensure that the separation between the impurities 
is much smaller in comparison to the dimension of the supercell.

All the electronic structure calculations reported in this work have been performed using self-consistent tight-binding linear muffin-tin orbital 
(TB-LMTO) method with the atomic sphere approximation (ASA)and the 
Combined Correction\cite{lmtoasa}.
TB-LMTO-ASA has been established as an intelligible, fast and accurate method
to understand electronic structure and chemical bonding for a large case of
solids, including systems with pronounced directional bonding\cite{oka,pawl}. Recently we have 
employed\cite{brkid} the TB-LMTO-ASA method coupled with the crystal orbital Hamiltonian population(COHP)\cite{cohp} for an energy-resolved visualization of chemical bonding in 
half-Heusler systems.
Self-consistent TB-LMTO-ASA calculations are done in the framework
of LDA\cite{lda}. We have also checked our calculations within
generalized gradient approximation (GGA)\cite{gga} and the results 
are found to be 
very similar to the LDA results.
The space filling in the ASA is achieved by inserting empty
spheres in the cube center positions
($\frac{3}{4}$,$\frac{3}{4}$,$\frac{3}{4}$)
and by inflating the atom-centered non-overlapping spheres. 
The atomic radii
are chosen in such a way that there is negligible charge on the empty spheres 
and the overlap of the interstitial with the interstitial
, atomic with atomic and interstitial with the atomic spheres remains within
the permissible limit of the ASA. The basis set of the self-consistent electronic structure calculation includes X,Y,M (s,p,d) and Z(s,p) and the rest are
downfolded. A [16,16,16] k-mesh has been used for self-consistency for 
all the calculations except for largest supercell where we have 
used [8,8,8] k-mesh. All the 
k-space integration was performed using the tetrahedron method\cite{tetra}.
Experimental lattice parameters 5.82$\AA$,5.88$\AA$\cite{evers},5.924$\AA$\cite{alliev}
for FeVSb, CoTiSb and NiTiSn respectively were used
for the doped structure and structural relaxations around the impurities 
were not included in the present calculations.
%%%%%%%%%%%%%%%%%%%%%%%%%%%%%%%%%%%%%%%%%%%%%%%%%%%%%%%%%%%%%%%%%%%%%%%%%
\begin{figure}
\begin{center}
\epsfig{figure=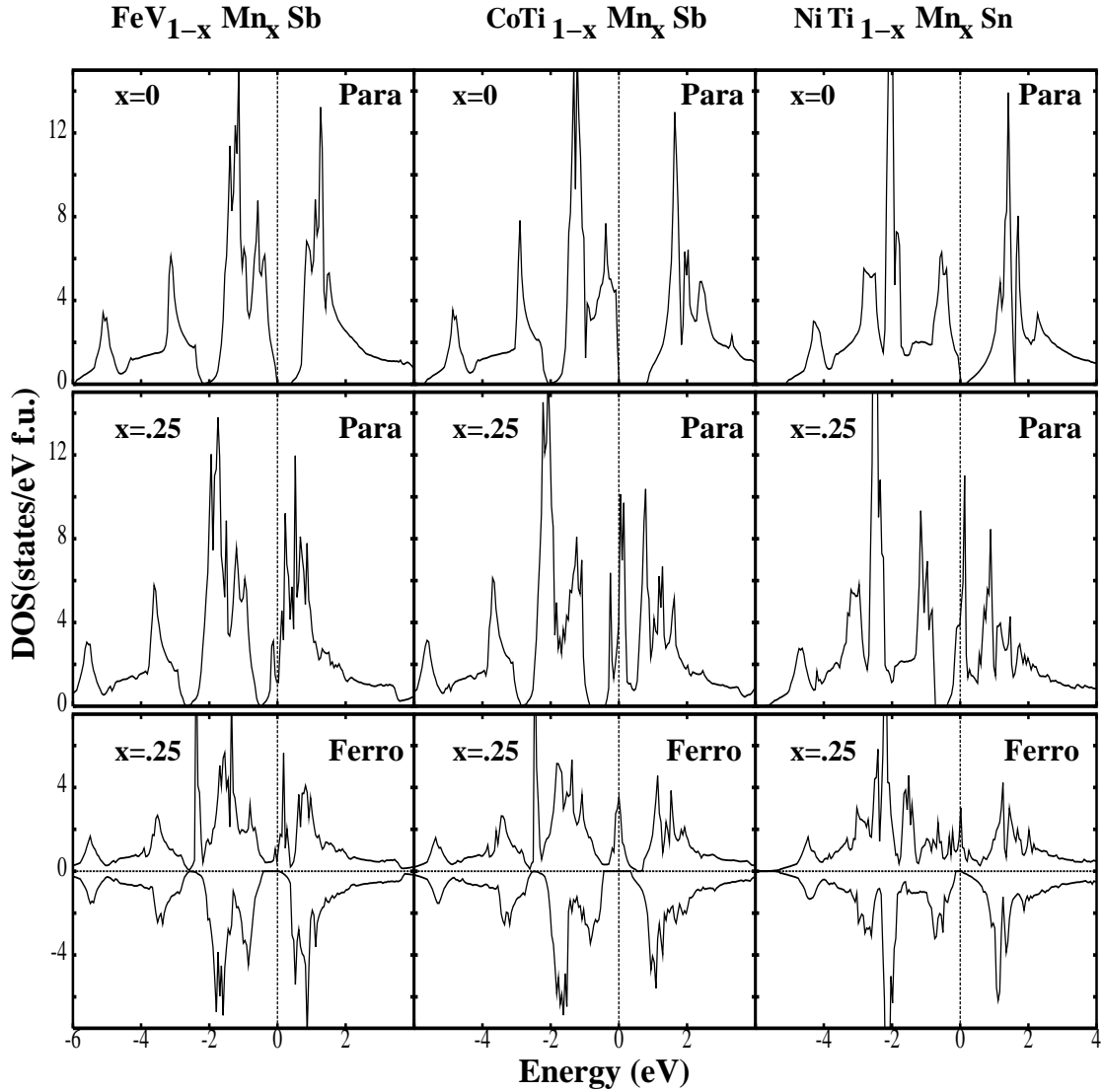,width=0.95\textwidth,height=0.62\textheight}
\caption{Density of states for formula unit for FeV$_{1-x}$Mn$_x$Sb(first panel), CoTi$_{1-x}
$Mn$_x$Sb (second panel)and NiTi$_{1-x}$Mn$_x$Sn(third panel). The first row shows the paramagnetic 
density of states of the semiconducting compound (x=0.0). Second and third row respectively shows the paramagnetic and ferromagnetic density of states of the Mn doped compound (x=0.25). All energies are w.r.t. Fermi energy.}
\end{center}
\end{figure}
%%%%%%%%%%%%%%%%%%%%%%%%%%%%%%%%%%%%%%%%%%%%%%%%%%%%%%%%%%%%%%%%%%%%%%%%%
\section{Results and discussions}
\subsection { Paramagnetic electronic structure of XM$_{x}$Y$_{1-x}$Z}
In this section, we shall present the results of the TB-LMTO-ASA 
electronic structure calculations for the Mn doped semiconducting 
half-Heusler systems in the unstable paramagnetic phase. We have considered 
three narrow gap semiconducting half-Heusler systems, FeVSb, CoTiSb and 
NiTiSn as the host material.
In each system the lower valent transition element (V and Ti) is 
substituted with M (Mn,Cr)
with concentration ranging from 25$\%$ to 3.125$\%$. 
Our calculations\cite{unpub} suggest that M substitution at other sites namely X(Fe,Co,Ni),
Z(Sn,Sb) and the voids is energetically unfavorable. 
This result is consistent 
with a recent KKR-CPA calculations
on some of these systems\cite{Tobola}. In first and second rows of figure 2 
first and second row we have displayed the total density of states(DOS) 
for the semiconducting half-Heusler hosts and the doped systems 
XMn$_{0.25}$Y$_{0.75}$Z in the paramagnetic phase respectively.
 Before we discuss 
the electronic structure of the doped systems we shall briefly discuss 
the electronic structure of the host materials. 

The characteristic 
feature of the electronic structure of the 
undoped compounds (XYZ, VEC=18, semiconducting) shown in 
the first row of figure 2 is a pair of bonding and antibonding states 
separated by a gap. The bonding states below the gap are predominantly of 
X character while the antibonding states are of Y character. The 
semiconducting gap is a consequence of the covalent hybridization 
of the higher valent transition element X with the 
lower valent transition element Y. 
Further the bonding states are well separated from the Sb-p 
states in FeVSb and CoTiSb by a small pd gap, while the Sb-s states 
are lying far below the chosen scale of the figure. In NiTiSn, the Sn p 
states lies higher in energy compared to Sb, 
overlapping with the bonding complex and therefore
the p-d gap does not exist. Again the Sn-s states are
far below the chosen scale of the figure. 
So below the d-d gap there are 9 bands 
(4 Z s+p, and 5 predominantly X-d) which in the paramagnetic phase can 
accommodate 18 electrons of both spins. Hence with 18 valence electrons all 
the s-p and d states below the d-d gap are saturated resulting in 
bond-orbitals with strong directionality and bonding. This explains 
why half-Heusler systems with 18 valence electrons are 
semiconductors\cite{brkid,dede}. 
We note that the presence of the Z(Sb/Sn)
atom which provides a channel to accommodate some transition metal d electrons in addition to its sp electron is therefore 
crucial for the stability of these systems.
For the half-Heusler systems with more than 18
valence electrons the antibonding states are occupied, and if the DOS
at the Fermi level is high then the paramagnetic state is no longer stable
and the stability can be achieved by developing a magnetic order. 
We have recently shown\cite{brkid} that the presence of a gap in the paramagnetic phase
promotes half-metallic ferromagnetism for systems with VEC $>$ 18 as for 
{\it e.g.} realized in NiMnSb.

We shall now consider the doped systems in the paramagnetic phase 
shown in the second row of figure 2.
In the doped systems replacing Ti/V with Mn puts an additional three/two electron 
in the system per Mn (VEC $>$ 18), as a consequence the systems are no longer 
semiconducting. This is reflected in the DOS for 25$\%$ Mn doped compound as a 
deep donor level produced by the addition of Mn impurities and the Fermi level 
lies on this predominantly Mn driven state. It is clear from the DOS peaks,
that the paramagnetic phase has a Stoner instability and a ferromagnetic state 
with finite moment is likely to be energetically favorable to the 
paramagnetic state.  The calculated DOS at the Fermi level $D(E_{F})$ 
is 2.21, 6.2 and 7.5 states/eV/Mn for 25 $\%$ Mn-doped FeVSb, CoTiSb and 
NiTiSn respectively. Assuming the usual value of the
Stoner interaction parameter $I_{Mn}$=0.75eV leads t0 $D(E_{f})I$
$\simeq$ 1.7, 4.6, 5.6, suggesting Stoner instability,
particularly strong for Mn doped NiTiSn and 
considerable energy gain via spin polarization.
\subsection{Spin-polarized Calculations}
%%%%%%%%%%%%%%%%%%%%%%%%%%%%%%%%%%%%%%%%%%%%%%%%%%%%%%%%%%%%%%%%%%%%%%%%%
% Put Fig. 3 here
\begin{figure}
\begin{center}
\epsfig{figure=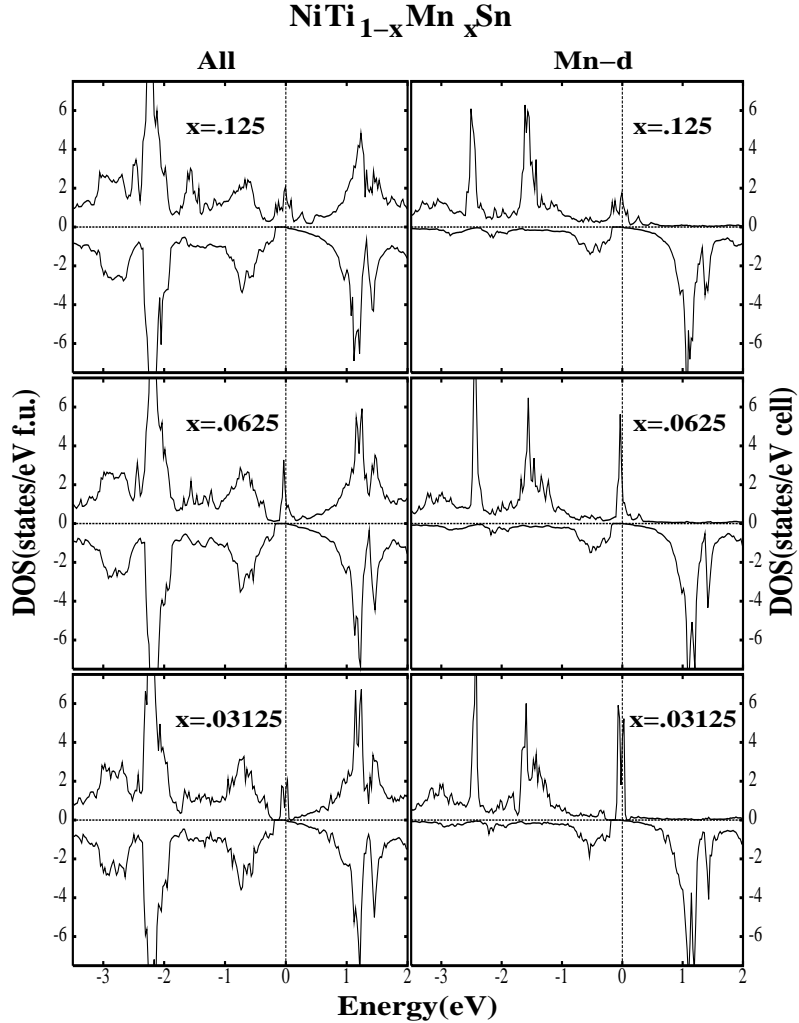,width=0.67\textwidth,height=0.58\textheight}
\caption{Spin-polarized density of states for NiTi$_{1-x}$Mn$_x$Sn (x=12.5$\%$, 6.25$\%$, 3.125$\%$). In the left column total density of states per formula unit are plotted while in the right column the Mn-d states per cell are plotted. All energies are w.r.t. Fermi energy.}
\end{center}
\end{figure}
%%%%%%%%%%%%%%%%%%%%%%%%%%%%%%%%%%%%%%%%%%%%%%%%%%%%%%%%%%%%%%%%%%%%%%%%%
Figure 2 (third row)
displays the spin-polarized DOS for the 25$\%$ Mn doped compound
in the semiconducting host. From the figure 
we gather that the doped systems are not only ferromagnetic 
but also half-metallic sustaining an integral moment 
of 3 $\mu_{B}$ for Mn in NiTiSn, CoTiSb and 2 $\mu_{B}$ in FeVSb(see table 1). 
Interestingly the defect states are screened metallically 
by the majority states so that the number of minority states do not change. The excess charge $\Delta z$ =3,3,2 per Mn in NiTiSn, CoTiSb and FeVSb
respectively are accommodated in the majority spin channel, resulting 
in electronic states appearing as a new peak in the gap, which are 
primarily of Mn character. Hence the charge which is 
excess of n $\times$ 18 (n = size of the supercell) appears as magnetic moment
satisfying the 18 electron rule suggested for half-Heusler systems with 
VEC $>$ 18\cite{dede}.

We shall now discuss in details
the role of diluted magnetic impurities (Mn and Cr) 
in the three semiconducting hosts. In each case the lower-valent transition element Y has been replaced with M(Mn/Cr) XM$_{x}$Y$_{1-x}$Z with x=12.5 $\%$,
6.25 $\%$, and 3.125 $\%$. For the purpose of discussion we have 
chosen a representative compound (NiTi$_{1-x}$Mn$_{x}$Sn), however 
our discussions holds for the other compounds as well. In figure 3 
we have displayed the total DOS for one formula unit 
and the site projected Mn DOS for concentration
x = 12.5 $\%$, 6.25 $\%$, and 3.125 $\%$. We gather from the total DOS 
that all the systems are half-metallic ferromagnets.
When Mn is substituted for Ti,
the additional 3 electrons per Mn are accommodated in the majority 
spin channel producing a Mn-derived states in the gap while the minority
states remain nearly unchanged.
The width of the impurity states, which is a consequence of the 
hybridization with the transition elements(Ni,Ti) and Sn, decreases
when the concentration of the impurity M is low. 
We also note that the Mn-derived states are partially filled.
Further we gather from the Mn partial DOS, there are sharp Mn derived crystal
field resonances deep in the valence band. 

A Ti vacancy in NiTiSn creates holes in the Ni-d-Sn-p valence band hybrid
due to depletion of four electrons per Ti, 
leaving the Ni-d Sn-p dangling bonds unsaturated.
 When Mn having seven valence electrons is substituted, 
four electrons are utilized to saturate the dangling bonds and the additional three electrons , responsible for the magnetic properties are accommodated in the high spin-state in the 
majority spin channel. This distribution of the Mn electrons results in crystal field resonances deep in the valence band, in addition to the partially filled impurity-derived states in the gap in the majority spin channel. 
%%%%%%%%%%%%%%%%%%%%%%%%%%%%%%%%%%%%%%%%%%%%%%%%%%%%%%%%%%%%%%%%%%%%%%%%%%%%
% Put Figure 4 (level diagram)
\begin{figure}
\begin{center}
\epsfig{figure=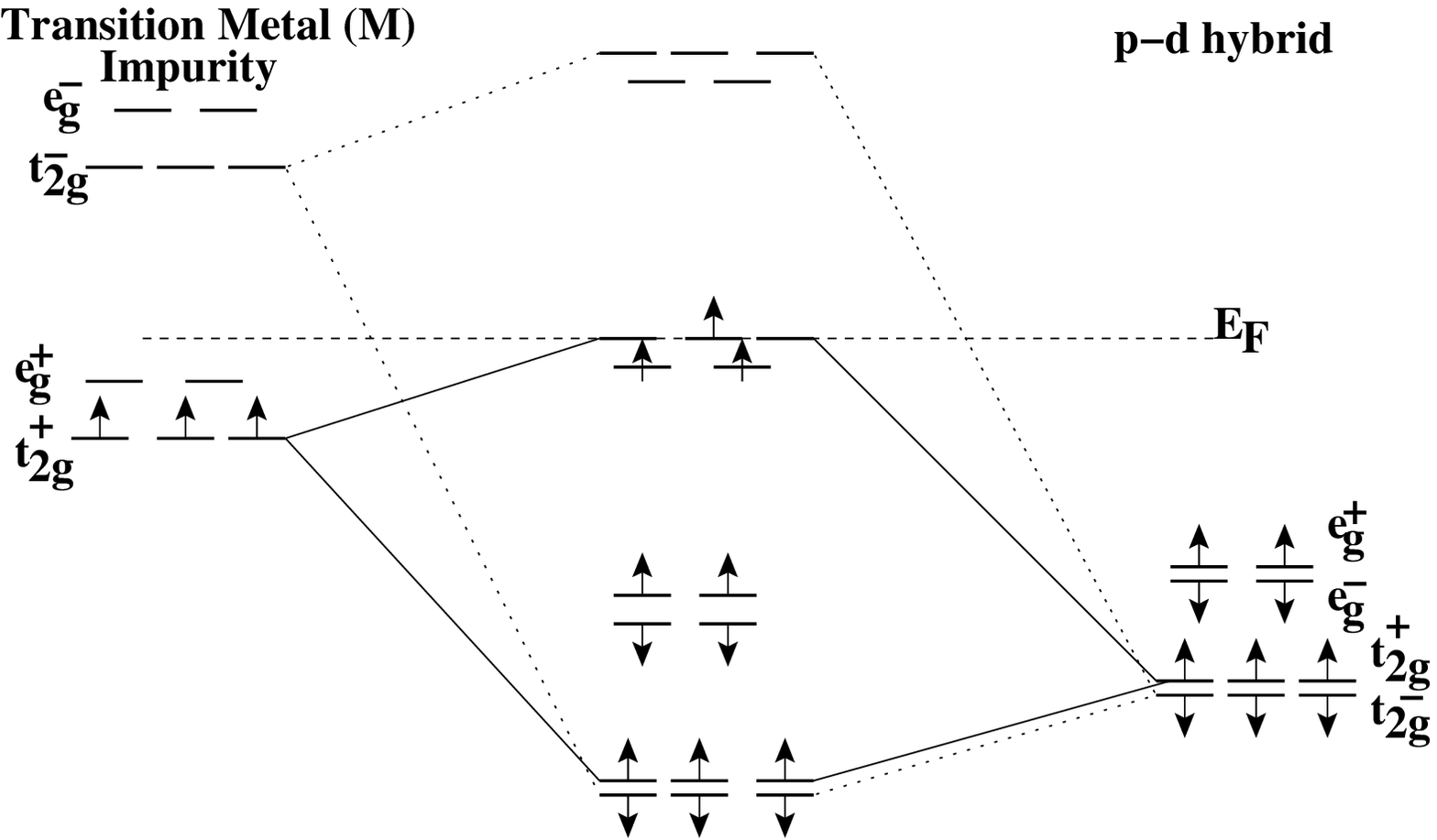,width=0.6\textwidth,height=0.37\textheight}
%\hfill
%\hspace{-1.0cm}
\epsfig{figure=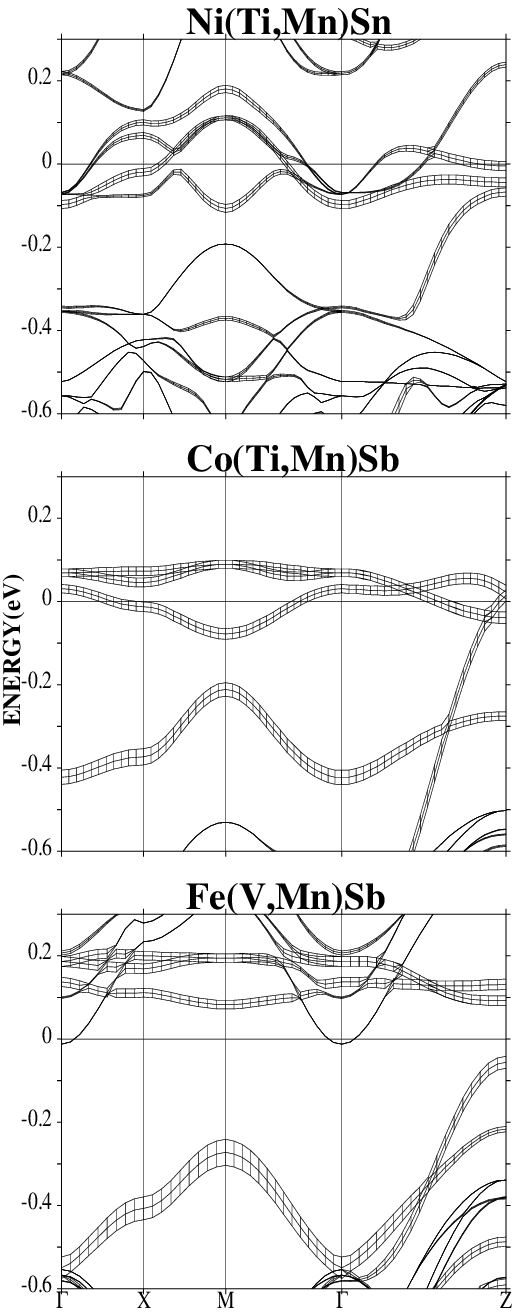,width=0.35\textwidth,height=0.37\textheight}
\caption{The gross feature of the electronic structure and genesis
of magnetism is shown through a energy level diagram in the left and in the right the Mn-d characters (fat bands) are plotted for Ni(Ti,Mn)Sn,Co(Ti,Mn)Sb and Fe(V,Mn)Sb with 6.25$\%$ impurity concentrations. All energies are w.r.t. Fermi energy.} 
\end{center}
\end{figure}
%%%%%%%%%%%%%%%%%%%%%%%%%%%%%%%%%%%%%%%%%%%%%%%%%%%%%%%%%%%%%%%%%%%%%%%%%%%

The gross feature of the electronic structure and magnetism 
in transition metal doped semiconducting 
half-Heusler compounds can be understood by invoking 
a simple model, where we consider the interaction between 
the dopant transition metal M and the valence band X-Z hybrid. Our simple model is schematically sketched in figure 4. In view of the strong 
d-d interaction, the valence band hybrid to a first approximation is primarily 
of X-d character, which
is split into t$_{2g}$ and e$_{g}$ levels by the crystal field.
After acquiring 
transition metal M electrons it is full and weakly spin split as illustrated 
in figure 4 (right panel). On the other hand the transition metal M-d levels 
are energetically shallower compared to the valence band hybrid. These levels 
are crystal field split as well as appreciably exchange split
, as shown in figure 4 (left panel). 
The result in the presence of hybridization is shown 
in the central panel of the figure. Since the X atoms are arranged tetrahedrally to the M atoms strong X-t$_{2g}$-M-t$_{2g}$ interaction is favored with 
relatively weaker e$_{g}$-e$_{g}$ interaction. The up(+) and down(-) spin states on the transition metal atoms therefore interact via spin-conserving hopping interactions and form a set of bonding-antibonding states for each spin channel, as shown in the central panel of figure 4. We note from the central panel
strong X-t$_{2g}$-M-t$_{2g}$ hybridization places transition metal M-t$_{2g}$ levels above the $e_{g}$ levels, and the relative positions of these 
levels is governed by the strength of the X-t$_{2g}$-M-t$_{2g}$ hybridization.
We note from the M projected band structure(fat bands) shown  
in figure 4(right) that the bands close to the Fermi level are partially filled Mn-d bands as expected.
We have recently shown\cite{brkid}
 with the aid of a tight-binding analysis that this hybridization depends on the relative position of the transition elements X and M 
in the periodic table . As a consequence Fe-M hybridization 
$>$ Co-M hybridization $>$ Ni-M hybridization. This is further illustrated in figure 5, where we have displayed M (Mn/Cr) projected partial DOS in a narrow 
energy range about the Fermi level for XM$_{x}$Y$_{1-x}$Z, 
with x=6.25$\%$(only the spin up channel is shown as the 
spin down channel is empty in this energy range).
We note from the figure that although the the $e_{g}$ and
t$_{2g}$ states are well split in Mn-doped FeVSb, the splitting decreases for CoTiSb, and is very small for Mn-doped NiTiSn, where the states are overlapping.
So for Mn doped 
NiTiSn the additional three electrons progressively fill up the overlapping
e$_{g}$ and t$_{2g}$ levels, with the t$_{2g}$ partially empty. 
The same is 
true for Mn doped CoTiSb, while the $e_{g}$ 
states are completely full for Mn-doped FeVSb (See figure 5).
For the Cr doped systems the splitting is very small for all semiconducting 
hosts leaving the Cr-d states partially empty.
%%%%%%%%%%%%%%%%%%%%%%%%%%%%%%%%%%%%%%%%%%%%%%%%%%%%%%%%%%%%%%%%%
% Put Fig. 5 here
\begin{figure}
\begin{center}
\epsfig{figure=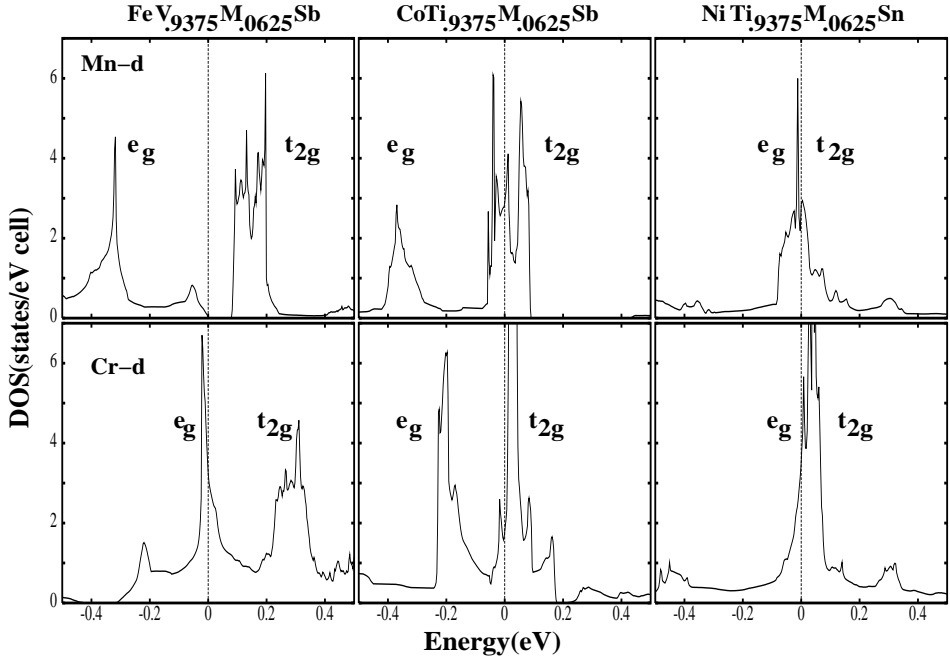,width=0.82\textwidth,height=0.38\textheight}
\caption{Spin-Up Mn-d(top) and Cr-d(bottom) density of states for the doped compounds with 6.25$\%$concentration. All energies are w.r.t. Fermi energy.}
\end{center}
\end{figure}
%%%%%%%%%%%%%%%%%%%%%%%%%%%%%%%%%%%%%%%%%%%%%%%%%%%%%%%%%%%%%%%%%%
%%%%%%%%%%%%%%%%%%%%%%%%%%%%%%%%%%%%%%%%%%%%%%%%%%%%%%%%%%%%%%%%%
% Put table 1 here
\begin{table}
\caption{Magnetic moments in $\mu_B$ of doped half-Heusler compounds}
%\begin{indented}
\begin{tabular}{@{}lllllll}
\br
Host&Impurity&Impurity&Total&M&p-d hybrid&Y+E\\
compound&M&Conc.&Mag.Mom.&Mag.Mom.&(X+Z)&Mag.Mom.\\
&&in $\%$&&&Mag.Mom.&\\
\mr
FeVSb&Mn&25.00&2.00&2.67&-0.98&0.31\\
&&12.50&2.00&2.64&-1.15&0.51\\
&&6.250&2.00&2.61&-1.34&0.73\\
&&3.125&2.00&2.58&-1.45&0.87\\
CoTiSb&&25.00&3.00&3.25&0.05&-0.30\\
&&12.50&3.00&3.24&-0.05&-0.19\\
&&6.250&3.00&3.24&-0.16&-0.08\\
&&3.125&3.00&3.22&-0.27&0.05\\
NiTiSn&&25.00&3.00&3.44&0.04&-0.48\\
&&12.50&3.00&3.41&0.07&-0.48\\
&&6.250&3.00&3.43&0.06&-0.49\\
&&3.125&3.00&3.40&0.01&-0.41\\
FeVSb&Cr&25.00&1.00&1.68&-0.95&0.27\\
&&12.50&1.00&1.54&-1.02&0.48\\
&&6.250&1.00&1.37&-1.12&0.75\\
CoTiSb&&25.00&2.00&2.58&-0.49&-0.09\\
&&12.50&2.00&2.53&-0.50&-0.03\\
&&6.250&2.00&2.56&-0.69&0.13\\
NiTiSn&&25.00&1.97&2.77&-0.25&-0.55\\
&&12.50&1.85&2.64&-0.23&-0.56\\
&&6.250&1.51&2.40&-0.25&-0.64\\
\br
\end{tabular}
%\end{indented}
\end{table}
%%%%%%%%%%%%%%%%%%%%%%%%%%%%%%%%%%%%%%%%%%%%%%%%%%%%%%%%%%%%%%%%%
The results of our calculations for the different 
hosts are summarized in table 1. 
We find from table 1 that the magnetic moment of M (Mn and Cr)
remain almost the 
same with the concentration variation of M, with the only 
exception being Cr-doped FeVSb. This insensitivity of 
magnetic moment with concentration is a signature 
of localized d sates of Mn and Cr. The localization comes from the 
fact that although d electrons of M are itinerant, 
the spin-down electrons are almost excluded from the M site.
Further in all cases (except Cr-doped NiTiSn) 
the total moment is always integer,
indicating a half-metallic solution. 

Finally, we have addressed the issue of the tendencies towards ferromagnetic 
or antiferroimagnetic ordering in these systems. For this purpose we 
have performed spin-polarized density functional calculations 
with (i) two M spins parallel to each other, the ferromagnetic configuration, 
(ii) two M spins anti-parallel to each other, 
the antiferromagnetic configuration. In order to understand the trends 
we have considered the hosts  NiTiSn, CoTiSb and FeVSb doped with Mn
and our results are 
summarized in table 2. 

%%%%%%%%%%%%%%%%%%%%%%%%%%%%%%%%%%%%%%%%%%%%%%%%%%%%%%%%%%%%%%%%%%
% Put table 2 here
\begin{table}
\caption{Magnetic Moments in $\mu_B$ and $\bigtriangleup$E(E$_{FM}$-E$_AFM$) for double impurity}
%\begin{indented}
\begin{tabular}{@{}llllllll}
\br
Host & Impurity& FM& & AFM& &$\bigtriangleup$E per Mn\\
compound& Mn conc.& Total  & Mn  &Total&Mn&meV\\
&&per Mn&&&&\\
\mr
FeVSb&25$\%$&2.00&2.64&0.00&2.65&-71.4\\
&12.5$\%$&2.00&2.60&0.00&2.57&-39.2\\
&6.25$\%$&2.00&2.56&0.00&2.56&-12.8\\
CoTiSb&25$\%$&3.00&3.22&0.00&3.28&-283.1\\
&12.5$\%$&3.00&3.21&0.00&3.22&-140.9\\
&6.25$\%$&3.00&3.21&0.00&3.16&-152.5\\
NiTiSn&25$\%$&3.00&3.36&0.00&3.48&-282.9\\
&12.5$\%$&3.00&3.36&0.00&3.43&-179.3\\
&6.25$\%$&3.00&3.32&0.00&3.38&-243.0\\
\br
\end{tabular}
%\end{indented}
\end{table}

%%%%%%%%%%%%%%%%%%%%%%%%%%%%%%%%%%%%%%%%%%%%%%%%%%%%%%%%%%%%%%%%

The stability of such a configuration in the context of diluted 
magnetic semiconductors is usually discussed in the framework of 
Anderson-Hasegawa model\cite{hasegawa}. If the M 3d orbital is partially occupied, 
in that case the 3d electrons in the partially occupied M 
orbitals 
are allowed to hop to the neighboring M 3-d orbitals, 
provided that the neighboring M ions have parallel magnetic moments. As a result, the d electrons lowers its kinetic energy by hopping in the ferromagnetic state.
This is the so called double exchange mechanism. However if the 
transition-metal M d orbitals are full then this reduction via 
hopping is not possible, and the energy is lowered by 
super-exchange which requires the neighboring spins to be anti-parallel.
In view of this it is expected that Mn-doped NiTiSn with partially occupied 
d levels should be ferromagnetic while Mn-doped FeVSb with empty 
t$_{2g}$ orbitals should favor antiferromagnetic arrangement of spins.
The total energy difference between FM and AFM configuration displayed 
in the last column of table 2 favors strong ferromagnetism for 
Mn doped NiTiSn and CoTiSb and is consistent with the Anderson-Hasegawa model,
while for Mn doped FeVSb instead of expected AFM, ferromagnetic state 
is found to be stable although the energy difference $\Delta E$
is very small. For the cases 
where the e$_{g}$ and t$_{2g}$ states are nearly overlapping
and the number of d electrons is less than five so that the
d-manifold is only partially occupied, ferromagnetism is always stabilized.
%%%%%%%%%%%%%%%%%%%%%%%%%%%%%%%%%%%%%%%%%%%%%%%%%%%%%%%%%%%%%%%%%
% Put Fig. 6 here
\begin{figure}
\begin{center}
\epsfig{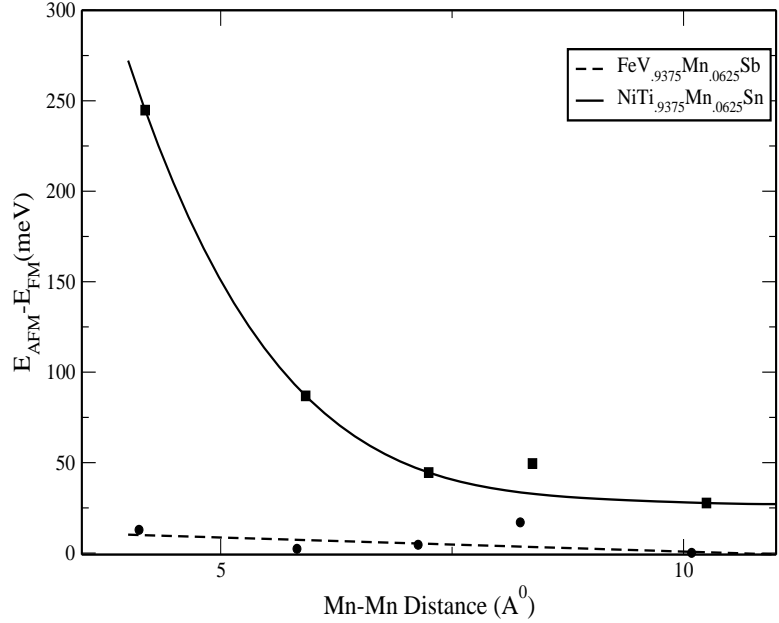}
\caption{Energy difference between the AFM and FM phase $\Delta E$ (= E$_{AFM}$-E$_{FM}$ w.r.t the Mn-Mn distance for FeV$_{.9375}$Mn$_{.0625}$Sb and NiTi$_{.9375}$Mn$_{.0625}$Sn.}
\end{center}
\end{figure}
%%%%%%%%%%%%%%%%%%%%%%%%%%%%%%%%%%%%%%%%%%%%%%%%%%%%%%%%%%%%%%%%%%
Finally, in figure 6 we have plotted $\Delta E$ as a function of 
Mn-Mn separation in the unit cell for Mn doped FeVSb and NiTiSn.
This difference in energy $\Delta E$ is also a measure of interatomic 
exchange interaction, and in the framework of mean-field theory is also 
proportional to T$_{c}$. Although $\Delta E$ is positive for both 
cases however it is very small for Mn doped FeVSb, suggestive of the fact 
that ferromagnetism may not be sustained in this compound. This 
is consistent with the Anderson-Hasegawa model. Appreciable 
$\Delta E$ for Mn doped NiTiSn suggests not only ferromagnetism 
for this compound but also appreciable T$_{c}$. Further, 
$\Delta E$ decreases sharply with distance suggesting that the 
double exchange mediated ferromagnetic interaction in these systems is short
ranged. A Similar finding has been reported for Mn doped GaN\cite{sanyal}, 
where it has been suggested short range interactions indicates the 
formation of Mn clusters within a short radial distance and might 
lead to high value of T$_{c}$ in these systems.
\section{Summary and Conclusions} 
We have studied in details the electronic structure and magnetism in 
Mn- and Cr- doped semiconducting half-Heusler systems, namely
FeVSb, CoTiSb and NiTiSn, for a wide concentration range. The 
characteristic feature of the electronic structure is a deep defect 
state of predominantly Mn/Cr character, in the majority spin channel.
The ferromagnetism in this systems can be understood within 
Anderson-Haswgawa model, which promotes ferromagnetism for partially 
filled d states. The electronic structure of the doped systems is analyzed
in the framework of a simple model where we have considered the interaction 
between the dopant transition metal M and the valence band X-Z hybrid.
We have shown that strong X-$t_{2g}$ and M-$t_{2g}$ interactions places 
the M -d sates close to the Fermi level, with the $t_{2g}$ states lying higher 
in energy in comparison to the $e_{g}$ sates. Further the splitting of the 
d levels is not only governed by the local crystal field but also
on the strength of hybridization between  X-$t_{2g}$ and M-$t_{2g}$ 
sates. Depending on the number of available 
d-electrons, ferromagnetism is realized provided the d-manifold is partially 
empty, as suggested by the Anderson-Hasegawa model. We have also addressed 
the issue of the tendencies toward ferromagnetic or antiferromagnetic 
ordering in these systems. Our calculations suggest that the double exchange mediated 
ferromagnetism in these systems is short ranged. Further the strong 
preference for ferromagnetic ordering over antiferromagnetic ordering 
in  Mn-doped NiTiSn indicates it  to be half-metallic 
ferromagnet with possibly high Curie temperature. Based on these theoretical 
predictions it will be interesting to investigate these systems experimentally.
\ack{We thank S.D. Mahanti for useful discussions. BRKN thanks CSIR, India, for
research fellowship(SRF). The research is funded by CSIR (No. 03(0931)/01/EMR-II)

\end{document}